\documentclass[sigconf]{acmart}

\AtBeginDocument{%
  \providecommand\BibTeX{{%
    \normalfont B\kern-0.5em{\scshape i\kern-0.25em b}\kern-0.8em\TeX}}}

\copyrightyear{2023}
\acmYear{2023}
\setcopyright{rightsretained}
\acmConference[CHI '23]{Proceedings of the 2023 CHI Conference on Human Factors in Computing Systems}{April 23--28, 2023}{Hamburg, Germany}
\acmBooktitle{Proceedings of the 2023 CHI Conference on Human Factors in Computing Systems (CHI '23), April 23--28, 2023, Hamburg, Germany}\acmDOI{10.1145/3544548.3581390}
\acmISBN{978-1-4503-9421-5/23/04}

\usepackage{xcolor}
\usepackage{soul}

\usepackage{color}
\usepackage{bm}

\usepackage{tikz}

\definecolor{CrossCodeOrange}{RGB}{236, 113, 0}
\definecolor{CrossCodeBlue}{RGB}{21, 144, 231}

\definecolor{CrossCodeOrangeDark}{RGB}{79, 38, 2}
\definecolor{CrossCodeBlueDark}{RGB}{0, 55, 94}

\definecolor{CrossCodeOrangeLight}{RGB}{255, 228, 204}
\definecolor{CrossCodeBlueLight}{RGB}{199, 228, 255}

\definecolor{CrossCodeDarkBlue}{RGB}{47, 106, 179}

\definecolor{CodeRedColor}{RGB}{220, 73, 88}
\definecolor{CodeBlueColor}{RGB}{47, 106, 179}
\definecolor{CodeGrayColor}{RGB}{83, 83, 89}

\definecolor{lightergray}{RGB}{250, 250, 250}
\definecolor{bordergray}{RGB}{175,175,175}

\newcommand\code[2][]{\tikz[overlay]\node[fill=lightergray,inner sep=1pt, anchor=text, rectangle, rounded corners=1mm, line width=0.1mm, draw=bordergray!50, #1] {\ttfamily \textcolor{CodeGrayColor}{#2}};\phantom{\ttfamily #2}}
\newcommand\codealt[2][]{\tikz[overlay]\node[fill=gray!20,inner sep=1pt, anchor=text, rectangle, rounded corners=1mm, #1] {\ttfamily {#2}};\phantom{\ttfamily #2}}

\newcommand\BLUB[2][]{\tikz[overlay]\node[fill=CrossCodeBlueLight,inner sep=1pt, anchor=text, rectangle, rounded corners=1mm,#1] {\color{CrossCodeBlueDark}{#2}};\phantom{#2}}
\newcommand\REDB[2][]{\tikz[overlay]\node[fill=CrossCodeOrangeLight,inner sep=1pt, anchor=text, rectangle, rounded corners=1mm,#1] {\color{CrossCodeOrangeDark}{#2}};\phantom{#2}}

\newcommand\CodeRed[1]{\textcolor{CodeRedColor}{#1}}
\newcommand\CodeBlue[1]{\textcolor{CodeBlueColor}{#1}}
\newcommand\CodeBlack[1]{\textcolor{black}{#1}}

\begin{document}

\title{CrossCode: Multi-level Visualization of Program Execution}



\author{Devamardeep Hayatpur}
\affiliation{%
  \institution{University of California, San Diego}
  \streetaddress{}
  \city{La Jolla}
  \state{California}
  \country{USA}}
\email{dshayatpur@ucsd.edu}

\author{Daniel Wigdor}
\affiliation{%
  \institution{University of Toronto}
  \streetaddress{}
  \city{Toronto}
  \country{Canada}}
\email{daniel@dgp.toronto.edu}

\author{Haijun Xia}
\affiliation{%
  \institution{University of California, San Diego}
  \streetaddress{}
  \city{La Jolla}
  \state{California}
  \country{USA}}
\email{haijunxia@ucsd.edu}

\renewcommand{\shortauthors}{Hayatpur, et al.}

\begin{abstract}
Program visualizations help to form useful mental models of how programs work, and to reason and debug code. But these visualizations exist at a fixed level of abstraction, e.g., line-by-line. In contrast, programmers switch between many levels of abstraction when inspecting program behavior. Based on results from a formative study of hand-designed program visualizations, we designed \textsc{CrossCode}, a web-based program visualization system for JavaScript that leverages structural cues in syntax, control flow, and data flow to aggregate and navigate program execution across multiple levels of abstraction. In an exploratory qualitative study with experts, we found that \textsc{CrossCode} enabled participants to maintain a strong sense of place in program execution, was conducive to explaining program behavior, and helped track changes and updates to the program state.
\end{abstract}



\begin{teaserfigure}
  \includegraphics[width=\textwidth]{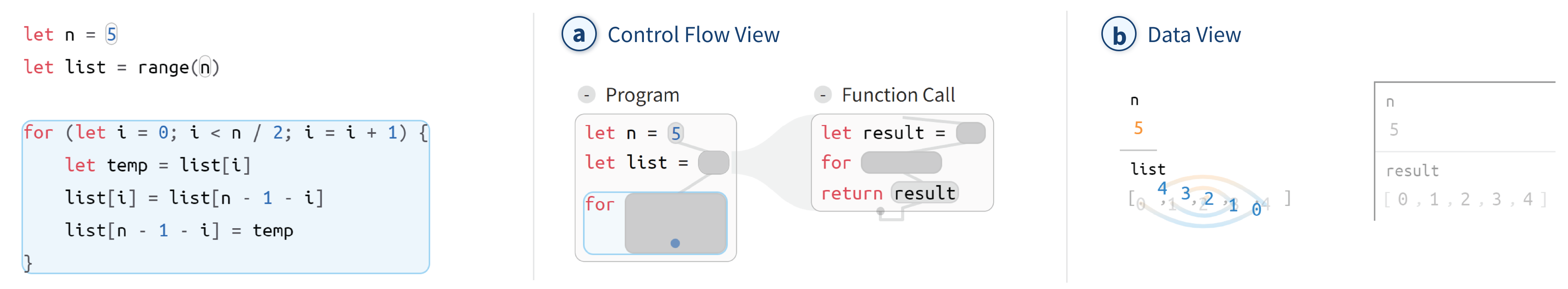}
  \caption{Screenshot from \textsc{CrossCode} for code to reverse a list. (a) An overview of the execution with which users interact to navigate levels of detail; (b) Program state and changes to data indicated with animations, traces, and color encoding.
  }
  \Description{A three column layout is shown, with textual source code on the left for an algorithm to reverse a list. In the middle is a miniature depiction of the source code, with control flow annotated on top, and one function-call is expanded but in a compact form. There is a blue cursor near the end of a for-loop. On the right is the program state, it shows the function-call's previous scope as well as an animation of items almost having been reversed.}
  \label{fig:teaser}
\end{teaserfigure}

\maketitle

\section{Introduction}
Computer programs translate human intent into machine-executable instructions. Their representations have largely remained the same: a set of strict textual instructions to generate a particular behavior. A programmer must translate their intent into precise instructions when writing code and then check execution details (e.g., observing an array update from \texttt{[5,1,2,3]} to \texttt{[1,5,3,2]}) for alignment with high-level intent (e.g., \textit{``sorting the first half of the array''}) when tracing and debugging code. Constantly ensuring congruence between the intended program behavior, the source code, and the execution demands excessive cognitive effort, becoming a learning barrier for beginners and a productivity impediment for experts.

Significant research has sought to reduce the cognitive effort of understanding computer programs by employing program visualizations. For example, tools like Python Tutor \cite{PythonTutor} and Projection Boxes \cite{ProjectionBoxes} display the runtime state of a program at each line. They enable programmers to answer common tracing questions like \textit{``What is the value of x at line 5?''}. However, they are less helpful in conveying overall program behavior, like \textit{``How does insertion-sort develop a sorted list?''}, which must be pieced together by stepping through the code line-by-line. In contrast to line-by-line navigation, algorithm visualizations use specialized representations to communicate the key steps and behaviors in the program \cite{AlgorithmVisualizationStateOfField, SalsaAlvis, Tango, MetaStudyAlgorithmEffectiveness}. However, these visualizations are not generically applicable: They need to be designed per algorithm or family of algorithms.

Therefore, existing systems either visualize \textit{low-level} program states, which are \textit{generically} applicable to many programs or represent \textit{higher-level} program behavior, which needs to be \textit{specifically} designed. In contrast, programmers and educators flexibly employ varying levels of abstraction to reason and communicate about programs. For example, a programmer may use breakpoints to inspect key program states and locate problematic sections, as well as hand-drawn diagrams to simulate program behavior and reason about program logic. Their reasoning processes can be bottom-up or top-down, depending on their expertise and tasks \cite{LETOVSKY1987}. Similarly, instructors often describe the same algorithm with gradually increasing levels of detail, starting with its purpose, then the main steps, and so on, moving between levels of abstraction. This research aims to fill the gap that exists in program visualization tools and can be summarized by the following problem statement: \textit{Given that programmers reason about code across multiple levels of abstractions, how can we design a program visualization system that can change its level of abstraction according to the programmers' needs?}

Our goal is thus to identify and apply techniques to visualize program execution at multiple levels of abstraction and design supporting interactions to navigate between these levels. To inform our design, we surveyed 92 computer science diagrams and animations from well-known instructional media for consistent visual communication strategies. We identified three key design patterns performed by these visualizations: they aggregated operations (e.g., instead of \textit{``x to temp to y,''} display \textit{``x to y''}), abbreviated repetitive operations (e.g., summarizing a loop by showing the first few and the last iterations), and displayed an overview of the execution space (e.g., depicting the call graph of a recursive function). 

Based on our findings, we developed \textsc{CrossCode}, a visualization system for a subset of JavaScript that takes advantage of the intrinsic features of a program, such as its syntactic structures, control flow, and data flow, to provide a flexible representation of its execution (Figure \ref{fig:teaser}). \textsc{CrossCode} enables navigation between multiple levels of detail. Rather than navigate the runtime execution line-by-line, users can flexibly navigate to the desired levels of abstraction through the syntax nodes of the source code. \textsc{CrossCode} uses an intermediate mapping akin to a \textit{`mental picture'} of the control flow to contextualize and navigate between levels of abstraction (Figure \ref{fig:teaser}a). The data state is rendered with animations, trace paths, and color encoding, calculated from the data flow (Figure \ref{fig:teaser}b). \textsc{CrossCode}\footnote{https://github.com/hayatpur/crosscode} is implemented with a custom interpreter for a subset of JavaScript written in TypeScript and instantiated as a web front-end. 

To evaluate the utility of \textsc{CrossCode} and its implications, we conducted an exploratory study with six expert programmers. We found that compared to \textsc{Python Tutor}, a line-by-line program visualization tool, participants were better oriented in the program's execution with \textsc{CrossCode}. Participants were also able to effectively navigate across repetitive operations and locate steps of interest. We contribute (a) identification of design patterns used to communicate program behavior, (b) a research prototype capable of generating and navigating program execution at varying levels of detail, and (c) results from a qualitative study, which shows that participants perceive \textsc{CrossCode} to be effective in locating errors and facilitating program understanding and communication.

\section{Related Work}
This research draws on several threads in cognitive accounts of programming, program visualization, and dynamic representations. 
\subsection{Program Understanding}
Researchers have proposed several cognitive models to describe how programmers develop an understanding of the source code \cite{CognitiveDesignElements}. These models suggest that programmers do not always parse source code line-by-line, but instead skip parts of code and regularly scan back and forth. Bottom-up models propose that programmers incrementally chunk statements into higher-level abstractions \cite{PENNINGTON1987295}. Brooks suggests that programmers start with an overall hypothesis of a program and incrementally refine it \cite{BROOKS1983543}. Soloway et al. suggested that expert programmers rarely process each instruction individually, but instead think in terms of schemas, i.e., chunks of instructions, that achieve a particular behavior \cite{Soloway1984}. Mayrhauser et al. found evidence that when programs lacked familiar cues or expectations were violated, a bottom-up process started from the details of the source code to incrementally synthesize an overall understanding \cite{Von1995}. Still others suggest that bottom-up and top-down processes are used depending on the user's expertise and task \cite{LETOVSKY1987}.


A programmer must also emulate the computer, i.e., trace code. Tracing forms the foundation to enable a programmer to effectively read, write, and debug code but is difficult to learn to do well \cite{TracingAndWriting}. In a study that observed the effects of tracing on working memory, Crichton et al. found two primary strategies: linear tracing following the control flow from the beginning to the end, or on-demand tracing, starting from the end with variables of interest and tracing back through the code that affected the variable \cite{RoleOfWorkingMemory}. Their findings highlight the need to adapt to user tracing strategies. For example, a person using an on-demand strategy would want to follow the data dependencies of a variable, while those using a linear strategy would want to follow the control flow.

Ko et al. observed that, when debugging, developers anchor their search with the execution of the program to look for symptoms of failure \cite{Whyline}. After determining the relevant code, developers used its incoming and outgoing connections to locate errors. However, developers frequently lost context during tasks and were not able to keep track of their explorations. They hypothesized that this may be due to showing the source code for the entire file at once rather than at a more appropriate granularity.

The different accounts of how programmers form mental maps of code, trace code, and debug code, exemplify the need for flexible representations of computer programs.

\subsection{Visualizations of Program Behavior}
Displaying program states is a standard method that is used to understand, debug, and communicate code. For example, Python Tutor \cite{PythonTutor} uses a memory model graphic with variables, pointers, and memory values, while Thonny \cite{Thonny} overlays the results of expressions onto the source code to ground symbolic code with runtime values. To trace values over time, Victor demonstrated a live programming interface that annotated the values of relevant variables at each line of source code \cite{InventingOnPrinciple}. Lerner generalized Victor's visualization using projection boxes, which select a subset of the full semantics of the program to display, providing a customizable view of runtime information \cite{ProjectionBoxes}. Other research has also investigated externalizing control flow. Reacher visualizes function calls as a graph to orient the user in debugging tasks \cite{VisualizingCallGraphs}, Debugger Canvas displays each function call in a spatial fragment \cite{DebuggerCanvas}, and Schematic Tables visualize a cross between decision tables and data flow \cite{SchematicTables}. 



Algorithm animations use specialized representations to communicate key behaviors of common algorithms. Starting with early animations like Sorting out sorting \cite{baecker1998sorting}, these visualizations fluidly transition between relevant consecutive states of a program. They help users understand how an algorithm works by focusing on specific key steps and invariants \cite{MetaStudyAlgorithmEffectiveness}. However, algorithm animations lack generalizability in their abstractions and visual encodings. Different visualizations must be configured and specified for different types of algorithms \cite{Tango}. Similarly, conceptual diagrams and animations are a staple in computer science education. Due to the high degree of customization conceptual visualizations need, designers need to create such visualizations on a case-by-case basis by using graphic design tools to manually produce diagrams, which is a challenging and labor-intensive process \cite{ConceptualDiagrams}. 

We aim to bridge the gap between generic program visualizations and specialized visualizations like algorithm animations and conceptual diagrams.




\subsection{Representing Multiple Levels of Abstraction}
This research builds on work that sought to support the mental processes of users through flexible representations \cite{WritLarge, LadderOfAbstraction}. For example, Victor explored the ladder of abstraction of data, procedures, and iterations from an automobile simulation that broadly represented the behaviors and patterns of a system \cite{LadderOfAbstraction}. The WritLarge system proposed flexible representational axes that allowed users to semantically, structurally, and temporally transition content representations to suit their needs rather than needing to conform to the single representation required by a user interface \cite{WritLarge}. 

Within the domain of programming, Suh et al.'s Coding Strip system used comic strips to provide a conceptual abstraction of program behavior and ease the burden of learning complex program executions \cite{ComicsToIntroduceReinforce}. Omnicode, on the other hand, enabled abstraction over time by visualizing values of a variable over its lifetime with scatter plots \cite{OmniCode}. Storey et al. proposed a top-down approach to algorithm animation, where information was gradually made more granular using steps that expanded it into more specific instructions. For example, a \textit{`QuickSort'} step could expand into \textit{`Divide'}, \textit{`Conquer'}, and \textit{`Combine'} \cite{TopDownAlgorithm}. (Their visualization and steps were pre-designed rather than generated procedurally.)

The present research builds on this work by generating different levels of abstraction of program execution for generic programs and by employing novel representation techniques that enable navigation and access to these levels of abstraction.

\section{Formative Study}

\begin{figure*}[h]
  \includegraphics[width=\textwidth]{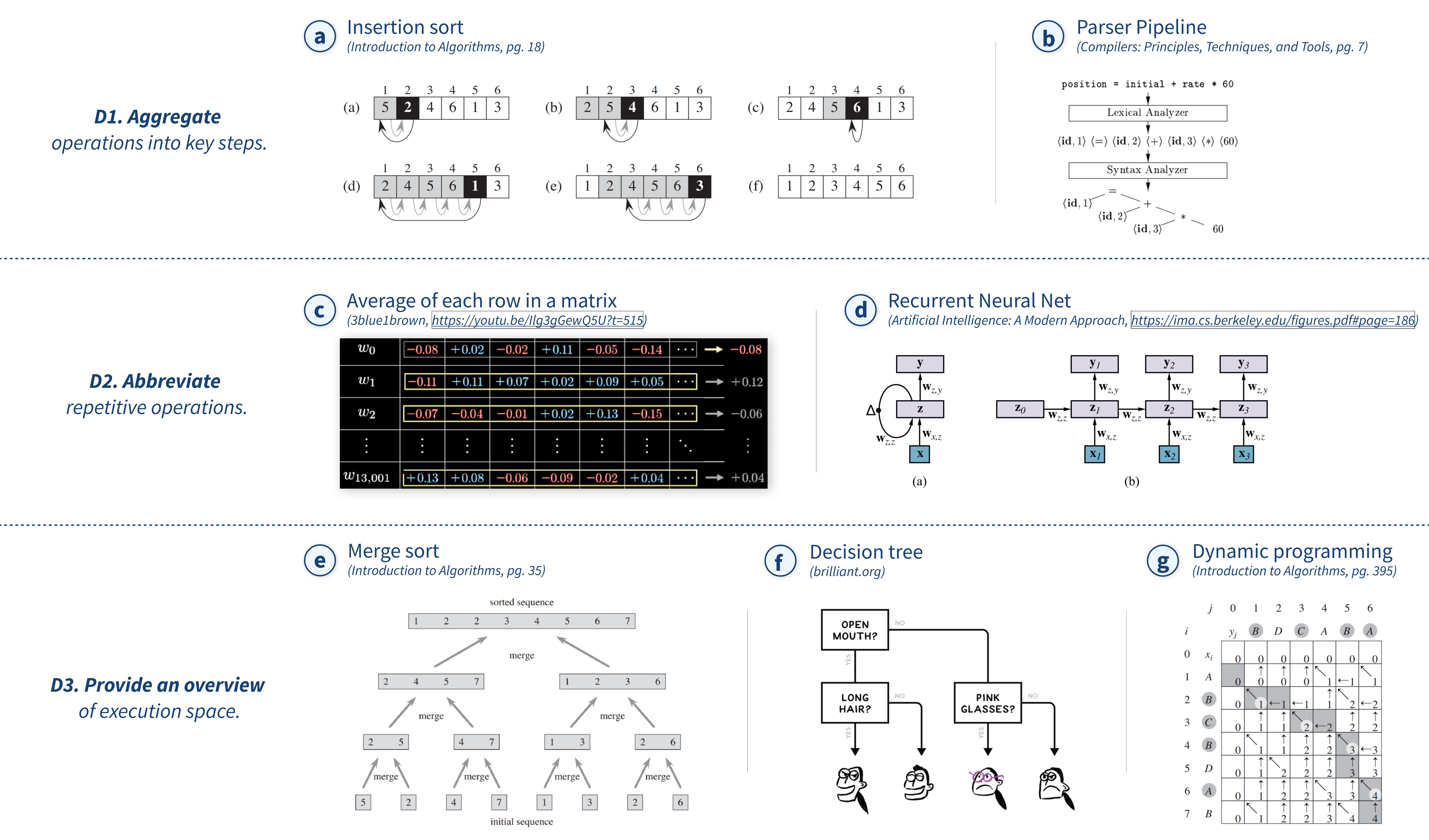}
  \caption{Examples of hand-designed illustrations that aggregated execution (a, b), abbreviate repetitive operations (c, d), and provide an overview of the execution space (e, f, g).}
  \Description{A grid of three rows, with multiple pictures in each row. The first row, from left to right, has: (a) a picture of six copies of a six item array, each copy shows array becoming sorted by inserting the proceeding item to the start; (b) a fragment of a parser which shows transformations to an initial input after two key steps (being, a lexical analyzer step, and then a syntax analyzer step). The second row, from left to right, has: (c) a picture showing each row of a matrix being summed, the first sum is completely visible, and the animation is in the process of summing the rest of the rows in simultaneously, the rows of the matrix are also abbreviated, only the first three and the last row are shown; (d) Two recurrent neural network depictions, one shows a neural network with its layers rolled up as a single loop, and the other shows each layer unrolled. The third row, from left to right, has (e) a merge-sort algorithm with the call graph shown in a tree structure, at the bottom is the base case shown as single item lists, and incrementally higher levels of the tree combine its children into a sorted sequence, (f) is a decision tree diagram, selecting from possible four possible cases after making two decisions, (g) is a dynamic programming table, the rows represent loop variable j and the columns represent the loop variable i, output of each call to the function is shown in its corresponding step.}
  \label{fig:handIllustrations}
\end{figure*}

To derive design patterns for flexible representations of program execution, handmade diagrams and animations used in computer science instructional materials were studied. These were selected as they are often transformed and simplified using various visual communication techniques, making them a suitable source of potential design patterns.

\subsection{Data Collection}

We sampled 92 visualizations from well-known textbooks, online instructional articles, and videos \cite{CLRS,ArtificialIntelligenceModern,NatureOfCode,CompilersPrinciplesTechniques,CraftingInterpreters,ProblemSolvingAlgorithms,StructureInterpretationOfComputerPrograms}, online instructional articles \cite{Brilliant, WikimediaCommons, OpenDSA} and videos \cite{3Blue1Brown, Reducible}. Some assumptions were made to constrain the types of visualization considered. First, if two visualizations used the same style and technique, then only one was used. Second, visualizations that directly matched program execution step by step and used no visual encodings were filtered out. This selection process resulted in 38 visualizations, 13 of which were animations. 

Each visualization was then annotated with low-level visual attributes by the first author, which described its design, e.g., \textit{``array represented as co-located boxes,''} \textit{``data flow represented with arrows,''} etc. The corresponding program source code described in the visualization was also annotated as it exemplified the deviations that the designer made from a potential implementation. The visual attributes and source code were then interpreted and grouped into broader design patterns to describe the process or transformation that could be used to communicate the program behaviors.

When deriving the design patterns, we followed Agrawala et al.'s visual design approach \cite{DesignPrinciples} and selected the following objectives each design pattern must satisfy: (a)\textit{ Program independence}, i.e., the pattern should generalize to a broad range of programs; (b) \textit{Generative}, i.e., there should be a well-defined criterion for how and when to apply the rule based on a provided source code and execution trace; (c) \textit{Cognitive rationale}, i.e., there should be a plausible hypothesis for how the rule leverages human perception and visual skills to aid in programming tasks.

\subsection{Design Patterns}


\subsubsection{D1: Aggregate Operations into Key Steps}


We found that designers communicated only key events of a program’s execution rather than all of its details. What constitutes a key event depends on the purpose of the visualization and the behavior described. In most cases, key events aggregated lower-level events based on the structure and semantics of the source code. For example, a useful invariant of insertion-sort is that elements are inserted at the start of the list. To illustrate this, operations in the inner loop of the algorithm, where individual elements are being swapped, were aggregated into a single movement (Figure \ref{fig:handIllustrations}a). Animation and visual annotations, such as arrows, depict the data flow to communicate the overall action of the step (Figure \ref{fig:handIllustrations}a). Aggregations can allow viewers to focus on program execution milestones without being overwhelmed by low-level steps (e.g., Figure \ref{fig:handIllustrations}b). When instantiating this pattern, we presume these milestones to be aggregations based on syntax; this assumption may not always be the case (i.e., the implementation may be separate from visual intuition), but it provides generalizability.

\subsubsection{D2: Abbreviate Repetetitive Operations}

Effort was made to reduce the amount of information for repetitive steps. For example, in loops with many iterations, only the first few and last are shown in detail, similar to how a large data structure is often abbreviated (e.g., writing a large list as [1, 2, ..., 13001]). This varies depending on the context, for example, in Figure \ref{fig:handIllustrations}c, only the first operation is shown in detail, and subsequent operations are aggregated into a single animation. Visualizations also exploited the symmetry of repeated operations with animations that sped up over time or diagrams that showed incrementally simpler representations. Abbreviations enable viewers to observe enough instances of a pattern to induct the general rule behind it, but without being overwhelmed by every instance of the pattern. We focus on two simple, but widespread repetitive structures when instantiating this pattern: linear loops and nested function-calls.

\subsubsection{D3: Provide an Overview of Execution Space}

An overview of the execution was often provided. For instance, execution of a recursive algorithm such as merge-sort algorithm is displayed through a call graph organized as a binary tree, where each node representing the state of the list (e.g., Figure \ref{fig:handIllustrations}e). Algorithms with double nested loops, such as for dynamic programming, often employed a grid, with one axis encoding iterations in the outer loop and the other encoding iterations of the inner loop (e.g., Figure \ref{fig:handIllustrations}g). In both cases, the control flow guides the layout, while the data flow and data values serve as landmarks: Figure \ref{fig:handIllustrations}e displays return value at each node, and \ref{fig:handIllustrations}g labels the values of the iterators on the axis. In other cases, the layout was guided mainly through the data flow, for example, a schematic of a neural network; where it better captures the underlying mathematical model (e.g., Figure \ref{fig:handIllustrations}d). An overview adds context to an execution step and enables visual deduction of overall patterns that would otherwise be difficult to synthesize in isolation. We choose to use the control flow to determine the layout, which can be readily mapped to multiple levels of abstraction.

\subsection{Other Patterns}
We uncovered patterns of data organization and encoding patterns that are not emphasized in our system, as they have been previously investigated and implemented, but are noted none-the-less:
\begin{enumerate}
    \item \textit{Variables are annotated around key data structures.} Visualizations preserved the relationships of different variables in a strong visual hierarchy. They used a focal data structure, with data and control structures annotated on top. For example, variables that index into an array was annotated on top of the array. Support for this customization exists in algorithm animation systems, e.g., Alvis Live \cite{SalsaAlvis, AlvisLive}, and customized placement of variables in Python Tutor \cite{PythonTutor}.
    \item \textit{Key data structures are temporally or spatially juxtaposed.} Data was frequently compared with itself or with other data by animations, providing an indication of change, or with small multiples to facilitate comparisons. Showing program states at multiple times has previously been explored by Omnicode \cite{OmniCode} and Projection Boxes \cite{ProjectionBoxes}, and animations by algorithm animation systems \cite{Tango, SalsaAlvis, AlvisLive}.
\end{enumerate}

\subsection{Relation to Prior Visual Techniques}
The three design patterns that we identified are not independent of prior research in visual communication techniques. In particular, Shneiderman's Visual Information-Seeking Mantra suggests that users should be able to (1) get an overview of the data, (2) zoom in on specific areas of interest, (3) filter the data to show only relevant information, and (4) access the details of the data on demand \cite{Mantra}. While Shneiderman originally intended these guidelines for understanding and interacting with complex data sets, our findings emphasize that the same underlying visual principles can be applied to understand and visualize code behavior. 

\begin{figure*}[h]
  \includegraphics[width=\textwidth]{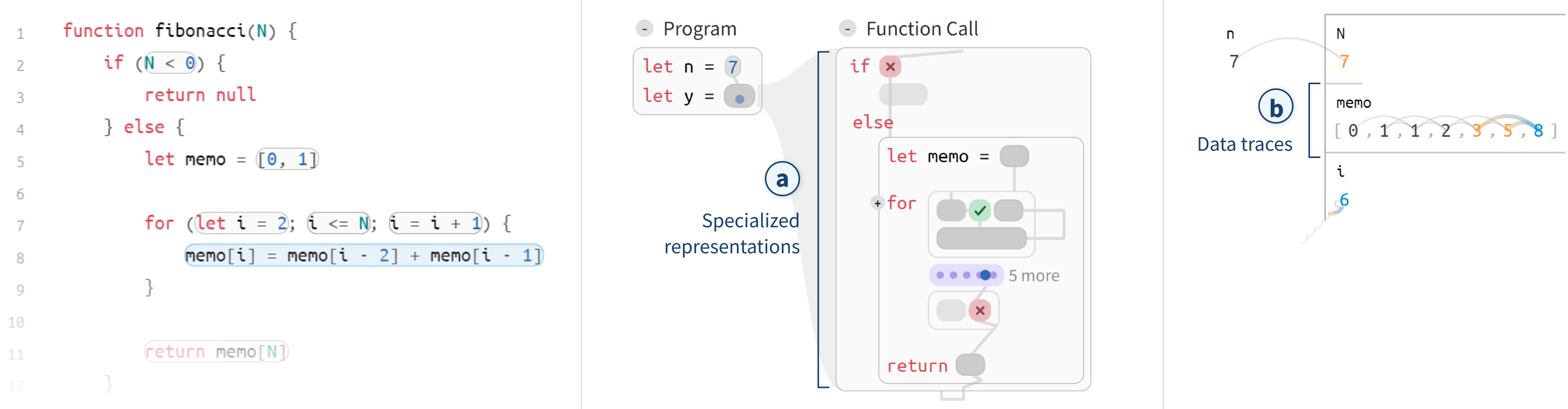}
  \caption{Execution of a Fibonacci algorithm. (a) Specialized representations are used for if-statements, for-loops, and function-calls; (b) Traces of data flow are rendered to help maintain a record of past \textit{Steps} and infer patterns over time.}
  \Description{Three column layout, with the left column containing source code mostly faded out. Middle, labelled (a), a miniaturized version the control flow. An unsuccessful if-statement explicitly indicates its execution through a red cross, a For Statement statement shows one iteration in full detail (a block for initialization, the test, the body, and the update), with subsequent iterations collapsed into dots. Right, (b), shows data which appears to be computing the Fibbinachi sequence using memoization, the final value in the array shows traces back to last two values which it is the sum of.}
  \label{fig:specialized}
\end{figure*}

\section{System Design}

Our formative study and prior work on understanding program behavior stress the need to view program execution at higher levels of abstraction. Here, we describe the design of \textsc{CrossCode}, which groups individual runtime steps using the code's syntax tree.

\subsection{Defining an Execution \textit{Step}}


The runtime of a JavaScript program can be modeled as registers, which store temporary values, and stack frames which map variable names to values. For example, \code{\CodeRed{let} \CodeBlack{x} = \CodeBlue{1} + \CodeBlue{2}} will first add \code{\CodeBlue{1} + \CodeBlue{2}} in a register, and then bind the output, \code{\CodeBlue{3}}, to \code{\CodeBlack{x}} in the current stack frame. A traditional program visualization may allow users to navigate through each of these steps one at a time, and view the memory state at each step. The key idea of \textsc{CrossCode} is to abstract sequences of individual steps, which are manipulation to registers and stack frames, into aggregate steps (or just \textit{Steps} for short). Specifically, we group individual steps based on the nodes of a program’s syntax tree. In doing so, we can query the execution based on the syntax, e.g., \textit{``What is the effect of this for-loop?''} Since the syntax tree is hierarchical, this grouping is also hierarchical, we can query for further details, e.g. \textit{``What is the effect of this if-statement in the for-loop?''}



For example, the loop: \code{\CodeRed{for} (\CodeRed{let} \CodeBlack{i} = \CodeBlue{0}; \CodeBlack{i} < \CodeBlue{1}; \CodeBlack{i}++) \{...\}} contains five sub-\textit{Steps}: (1) initialize: \code{\CodeRed{let} \CodeBlack{i} = \CodeBlue{0}}; (2) the first test: \code{\CodeBlack{i} < \CodeBlue{1}}, which succeeds; (3) the body: \code{\{...\}}; (4) the update: \code{\CodeBlack{i}++}; and (5) the second test: \code{\CodeBlack{i} < \CodeBlue{1}}, which fails and terminates the loop. These sub-\textit{Steps} can be decomposed further, e.g., the test: \code{\CodeBlack{i} < \CodeBlue{1}} is composed of two sub-\textit{Steps}: (1) the identifier \code{\CodeBlack{i}}, which reads from the stack frame and writes the value of \code{\CodeBlack{i}} to a register, it cannot be decomposed further; and (2) the literal \code{\CodeBlue{1}}, which writes the value \code{\CodeBlue{1}} to a register and cannot be decomposed further.

Our goals are to design (a) an understandable visualization of a \textit{Step} and (b) facilitate navigation across \textit{Steps}, both over breadth, such as navigating to a different iteration of a loop, and over depth, such as navigating across a stack of a recursive function calls. We achieve these goals through three views: the \textit{Control Flow View}, which visualizes the \textit{Steps} themselves, the \textit{Data View}, which shows the effect of the \textit{Step} on the program state, and \textit{Source Code View}, which grounds the \textit{Step} in the source code.


\subsection{Control Flow View}
\textsc{CrossCode} visualizes a \textit{Step} as a miniaturized block of source code from which it originated (Figure \ref{fig:specialized}). This representation includes fragments of the corresponding source code which serve as landmarks. For example, a variable-declaration will partially reveal its left-hand side (e.g., \code{\CodeRed{let} \CodeBlack{x} = \LARGE\textcolor{gray!70}{$\blacksquare$}}\normalsize). \textit{Steps} are organized according to their layout in the source code. The landmarks and consistent organization enable users to situate themselves in the control flow using their mental image of the source code.

The path control flow takes is annotated between each \textit{Step} and visually connects adjacent \textit{Steps}. A blue cursor on the path indicates where the program execution is currently (Figure \ref{fig:specialized}a). Users can navigate the program execution with keyboard controls (that is, stepping with left and right arrow keys), by directly grabbing and moving the cursor, or by clicking on a \textit{Step} to navigate to its end point. To navigate between levels of abstraction, a user can control-click on a \textit{Step} to break it down to a lower level of abstraction (i.e., its sub-\textit{Steps)}. A control-click on an already broken-down step will aggregate it back to a higher level of abstraction.

\begin{figure*}[h]
  \includegraphics[width=\textwidth]{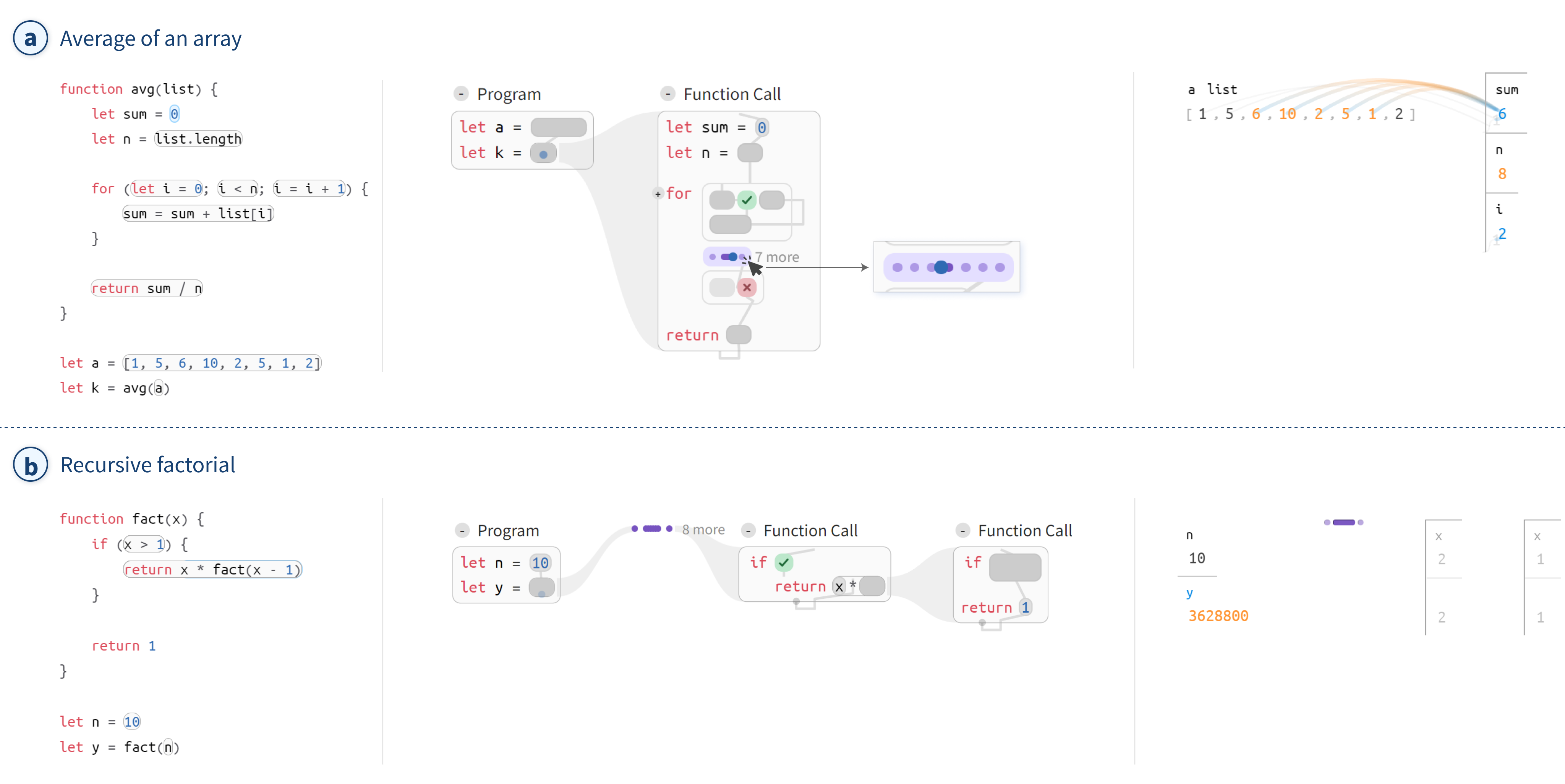}
  \caption{Abbreviations can be used for both loops and function-calls. (a) A program to compute the average of an array, subsequent iterations of the for-loop are abbreviated and their operations are aggregated, clicking on the abbreviation expands it to reveal a dot for each iteration individually. (b) A program to compute Factorial of \textit{n}, the user has navigated to the base case of the function, and parent function-calls are abbreviated.}
  \Description{Two rows. The top row, (a), shows miniaturized control flow for a function to sum items in the list, subsequent iterations of a for-loop are abbreviated into three dots, the cursor is on the middle dot. Clicking on the abbreviation reveals each iteration as its own dot, resulting in a group of 7 dots. There is a data execution to the right which shows the middle dot executing five iterations of the loop in aggregate. The bottom row, (b), shows miniaturized control flow for a Factorial of 10. It is currently at the base case but the call stack is abbreviated such that only the last two function-calls are shown. To the right, the scopes of function-calls are also correspondingly abbreviated.}
  \label{fig:abbreviation}
\end{figure*}

\subsubsection{Specialized Representations}
Specialized representations of if-statements, for-loops, and function-calls are used to better reflect each structure's meaning. For example, if-statements and for-loops indicate the result of their condition directly in the \textit{Control Flow View} using a checkmark for success or a cross for failure (Figure \ref{fig:specialized}a). They also contain stubs for \textit{Steps} that were not reached but exist in the source code (such as an unsuccessful body of an if-statement, for example, Figure \ref{fig:specialized}a), which help maintain a close match between the source code and the control flow, even when parts of the source code are not reached. Unlike other control flow structures, a function-call is not expanded in place, but rather in a separate space, termed a \textit{Frame} (Figure \ref{fig:specialized}a). This reflects that each function call exists in its stack frame, analogous to existing call graphs visualizations \cite{DebuggerCanvas, VisualizingCallGraphs}. Each \textit{Frame} has its own control flow cursor, which is synchronized across the other \textit{Frames}. This helps situate oneself relative to other function-calls, such as to know if the execution is currently before, after, or during the \textit{Frame}.

The \textit{Control Flow} View works towards D1 by enabling users to select the appropriate level of abstraction, and D3 by providing an overview that orients and guides the user's navigation. 

\subsection{Data View}
\textsc{CrossCode} externalizes the runtime state of a program through a view of the underlying memory. It displays columns of memory values with variable reference(s) to the value denoted above (Figure \ref{fig:specialized}b). Data in different stack frames are spatially separated, corresponding to the position of their associated function-call \textit{Frame} in the \textit{Control Flow View} (Figure \ref{fig:specialized}b).

\subsubsection{Animations and Traces.} To visualize transformations of data at multiple levels of abstraction, \textsc{CrossCode} derives an animation of a \textit{Step} using its reads and writes. Animations not only provide a mechanism to maintain context when following execution but also to show aggregate data transformations (e.g., Figure \ref{fig:specialized}b). A data \REDB{read} is colored orange, and a data \BLUB{write} is colored blue, providing a lookup of how the current step modified the state. We employ the following animation types ($\mathtt{\textbf{A}}$, $\mathtt{\textbf{B}}$, and $\mathtt{\textbf{X}_i}$ denote memory values):

\begin{enumerate}
    \item $Create$(\BLUB{$\bm{A}$}) is used when \BLUB{$\bm{A}$} is written to with no reads from other existing visualized data. A fade-in animation for \BLUB{$\bm{A}$} is used to represent it.
    \item $Move$(\REDB{$\bm{B}$} $\rightarrow$ \BLUB{$\bm{A}$}) is used when value \REDB{$\bm{B}$} is copied to value \BLUB{$\bm{A}$}. The resulting animation is shown as \BLUB{$\bm{A}$} originating from \REDB{$\bm{B}$} and gradually moving to its memory position. In addition, a trace is shown from data \BLUB{$\bm{A}$} to data \REDB{$\bm{B}$}.
    \item $Cause$(\REDB{$\bm{X}_0 ... \bm{X}_N$} $\rightarrow$ \BLUB{$\bm{A}$}) occur when data is written to \BLUB{$\bm{A}$} from multiple reads, \REDB{$\bm{X}_0 ... \bm{X}_N$}, (e.g., in a binary expression). In this case, a trace is shown from each read, \REDB{$\bm{X}_0 ... \textbf{X}_N$}, to \BLUB{$\bm{A}$}, along with a fade-in animation for \BLUB{$\bm{A}$}.
\end{enumerate}

Animations serve to highlight the change between data states, whereas traces serve as a persistent visualization of the operation. Traces of past executions are displayed but faded, allowing users to infer and abstract over time while maintaining visual clarity (e.g., Figure \ref{fig:specialized}b). Note that these animations cover a subset of possible data dependencies. Control flow dependencies such as conditional dependencies are not visualized (e.g., if value $\mathtt{\textbf{B}}$ being \verb|true| causes value $\mathtt{\textbf{A}}$ to change, that dependency is not recorded).

\subsubsection{Residuals}
Data that is replaced (e.g., \code{\CodeRed{let} \CodeBlack{i} = \CodeBlue{1}; \CodeBlack{i} = \CodeBlue{2}}) is pushed back by fading it and giving it a positional offset, analogous to crossing out an older value when tracing a program with pen and paper. Residuals make it possible to visualize traces of data that no longer exist in the program state without adding visual noise. For example, in an in-place swapping algorithm (Figure \ref{fig:teaser}b), the residuals disambiguate the origin of the traces. \\

The \textit{Data View} instantiates D1 by indicating aggregated steps through animations and data traces.

\subsection{Source Code View}
The source code is synchronized with \textit{Control Flow View}, in that the current \textit{Steps} being executed is annotated on top and highlighted. So far, \textsc{CrossCode}'s functionality has supported top-down methods of navigating abstractions, by starting from an overall \textit{Step} and incrementally breaking it down. The source code provides a map to freely navigate across levels of abstraction. When selecting any specific piece of the source code, the execution jumps to points at that level of abstraction. For example, a user can select a base case of a function-call to navigate to all points in the execution where that base case is executed.

\subsection{Supporting Multiple Levels of Abstractions}

Computer programs contain fundamental repetitive structures that can be impractical to show in their entirety. Rendering many iterations of a loop or displaying the entire control flow graph of a deeply nested function-call can easily become cluttered and difficult to reason about. \textsc{CrossCode} provides two mechanisms to manage this complexity: abbreviations and aggregations. Abbreviations instantiate D2: they provide structure to navigate and interpret repetitive procedures, and aggregations provide a zoomed-out overview of longer executions \textit{Steps}, further facilitating D3.

\begin{figure}[h]
  \includegraphics[width=0.5\textwidth]{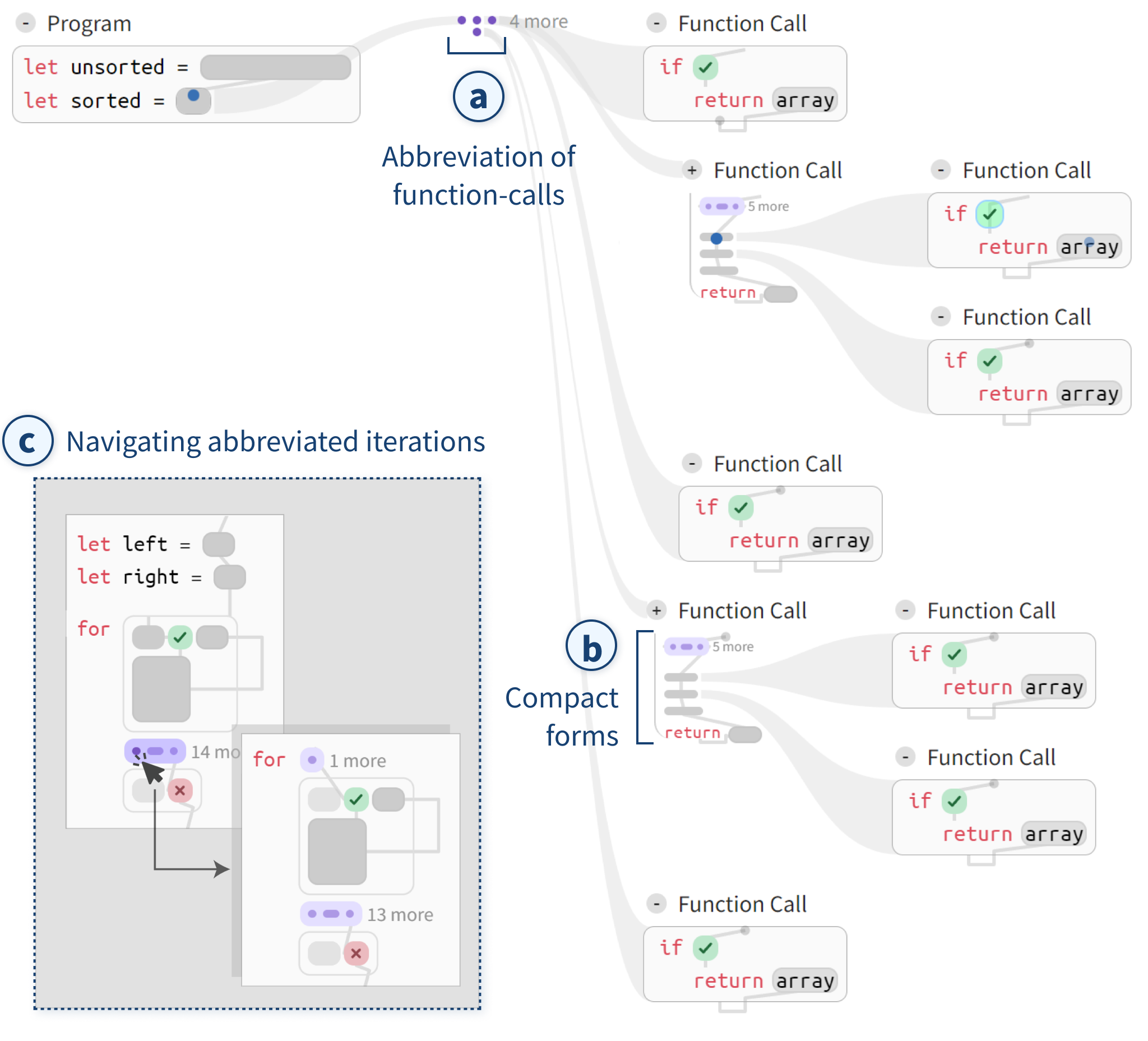}
  \caption{Execution of a recursive call to a Quicksort algorithm, expanded up till the base cases. (a) Progressive closure automatically abbreviates older function-calls; (b) Progressive disclosure only shows the last few steps of the Function, instead of showing all its sub-\textit{Steps}. (c) Control-clicking on an abbreviated iteration expands it and progressive closure abbreviates the previously expanded iteration. }
  \Description{In the background is a partially abbreviated call graph showing the base case of a recursive call to a quick-sort algorithm. It shows two visual techniques: (a), ancestor calls are collapsed to a dot, (b) parent calls are presented in a compact form, with the first few steps inside the parent function abbreviated. In the foreground is miniaturized version of a for-loop, it shows that by Control-clicking on an abbreviated iteration, it will expand that iteration into sub-steps and collapse the previous iteration that was expanded.}
  \label{fig:disclosure}
\end{figure}


\subsubsection{Abbreviations}

Abbreviations aim to trim the amount of repetitive information by collapsing a \textit{Step} into a dot, potentially grouped together with other dots (Figure \ref{fig:abbreviation}a). By default, an abbreviation of $\mathbf{N}$ sub-\textit{Steps} collapses its contained actions into three dots: the first represents sub-\textit{Step} $\mathbf{0}$, the second dot represents sub-\textit{Steps} $\mathbf{1}$ to $\mathbf{N - 1}$, and the last represents \textit{Step} $\mathbf{N}$. To revert an abbreviated \textit{Step} to its normal level of detail, the user can control-click on it (Figure \ref{fig:disclosure}c). Clicking on an abbreviation group toggles it to an non-aggregated mode, where each sub-\textit{Step} in the abbreviation is represented as a separate dot, that is, for $\mathbf{N}$ sub-\textit{Steps}, there will be $\mathbf{N}$ dots (Figure \ref{fig:abbreviation}a, middle). We also use specialized abbreviations for for-loops and function-calls:
\begin{itemize}
    \item For-loops do not abbreviate the initialization, the test, the body, and the update as separate dots, but instead aggregate them into iterations, i.e., a dot represents an entire iteration. By default, the first iteration is unabbreviated, and the rest of the iterations are abbreviated. In this way, users can view the first iteration in-depth and skim over subsequent iterations. 
    \item Function-calls that are nested can be abbreviated (Figure \ref{fig:disclosure}a), which also abbreviates the function-call's corresponding scope in the \textit{Data View} (Figure \ref{fig:abbreviation}b). This makes it feasible to browse deeply nested structures, such as recursive function calls, by an overview of the call stack without viewing each function-call in isolation. For example, in a quick-sort algorithm, a user can easily read the path from the parent call to the base case (Figure \ref{fig:disclosure}a).
\end{itemize}

\subsubsection{Compact Forms}

In addition, we provide a toggle to view \textit{Steps} in a compact form (Figure \ref{fig:disclosure}b). In their compact form, the landmarks, spacing, and overall correspondence of \textit{Step} to the source code are minimized in favor of a compressed representation.


\subsubsection{Progressive Disclosure and Closure (D3)}
Thus far, the two mechanisms to manage complexity, abbreviations, and compact forms, need to be applied manually. To help users maintain and establish context, we provide two modes in which these techniques are proactively applied.

\begin{enumerate}
    \item \textit{Progressive disclosure} aims to show new information incrementally rather than presenting all sub-\textit{Steps} at once. It does so by (a) automatically abbreviating for-loops, and (b) only showing the last four steps of a function-call body (Figure \ref{fig:disclosure}b). Note that if the user would like to see all iterations of a for-loop, we provide a toggle to show all the iterations of the loop, i.e. unroll the loop.
    \item \textit{Progressive closure} collapses older \textit{Steps} when a user expands a new \textit{Step}. For example, given \code{\CodeRed{f}(\CodeRed{g}(\CodeRed{h}(\CodeBlack{x})))}, if the current level of abstraction is at \code{\CodeRed{f}(...)}, then when a user expands to the first child \code{\CodeRed{g}}, \textsc{CrossCode} automatically sets the parent, \code{\CodeRed{f}}, to be in a compact representation (Figure \ref{fig:disclosure}b). If the user expands to the second child \code{\CodeRed{h}}, it abbreviates the ancestor \code{\CodeRed{f}} entirely (Figure \ref{fig:disclosure}a). In doing so, this two-stage collapsing can help organize space while retaining context. This helps navigate recursive functions or complex call stacks (Figure \ref{fig:disclosure}a,b). Progressive closure also automatically closes older iterations of a loop as the user un-abbreviates new ones (Figure \ref{fig:disclosure}c).
\end{enumerate}

In sum, \textsc{CrossCode} uses abbreviations and aggregations to manage complexity in larger executions. It invokes these techniques through progressive closure and disclosure of information, which we hope allow for effective navigation and viewing of multiple levels of abstraction.

\subsection{Implementation}
\textsc{CrossCode} is implemented with a custom interpreter and a web front-end written in TypeScript. For tooling, \textit{Vite} is used to compile and bundle the project \cite{vite}, and the \textit{Monaco Editor} \cite{vite} is used in the web front-end for the user to input source code. Herin, we describe the information pipeline:

\subsubsection{Dynamic Analysis} 
The user's source code is parsed into an abstract syntax tree using \textit{acorn} \cite{acorn}. A custom interpreter walks through the syntax tree, executing on a stack-based, virtual memory model, and returns a structured trace of the execution which groups the operations based on the syntax tree. Our interpreter architecture was based on \textit{Sval}, an open-source JavaScript interpreter \cite{sval}. Each node in the trace eventually decomposes into primitive operations, such as \codealt{CreateLiteral} (for creating a literal value in memory), \codealt{BinaryExpression} (for computing a binary expression), etc. Each primitive operation stores its reads and writes which are used to generate the data flow. In addition, all nodes store a precondition and a postcondition of the memory model before and after the execution of that node.

\subsubsection{Rendering \textit{Control Flow View} and \textit{Data View}} 
The structured execution trace derived from the dynamic analysis drives \textit{Control Flow View}. Each node in the trace is assigned a visual representation based on its type (e.g., a for-statement is assigned a different renderer than an if-statement). The visual representation is built and displayed through the DOM. Subsequently, the \textit{Control Flow View} drives the \textit{Data View}: based on the current node selected, the \textit{Data View} renders its postcondition, and based on the position of the cursor, it interpolates an animation derived from the data traces. The data traces themselves are rendered using the \textit{Perfect Arrows} library \cite{perfectarrows}. Users can edit the animation speed through an options panel rendered with \textit{Tweakpane} \cite{tweakpane}.

\subsection{Implementation Limitations}
While the techniques in \textsc{CrossCode} can likely generalize to other imperative languages, we focus on a single language: JavaScript. Specifically, the subset of JavaScript with constants, function calls and definitions, binary operators, let-bindings, if-statements, and loops. Notable exceptions include minimal support for objects, no classes or prototype inheritance, no implementation of standard global objects outside of a limited subset of \codealt{Math} object, no exception handling, and no support for DOM manipulation. (See Section \ref{section:Scalability} and \ref{section:Asynchrony} for a discussion of ideas for supporting classes and synchrony respectively.) 
 
A limitation to the adoption and generalizability of our current implementation is the use of a custom interpreter. Note that this is not a requirement: the information flows we use (i.e. control flow and data reads and writes) may be extracted reliably by instrumenting an existing runtime, such as of the browser, or by using a dynamic analysis library, e.g., Jalangi \cite{Jalangi}. We opted to use a custom interpreter for its flexibility and ease of use in accessing the information we required during prototyping.

Finally, not all runtime operations are externalized in \textsc{CrossCode}. For example, operations to create or pop scope are not made visible to the user as they are often aligned with another \textit{Step} (e.g. a for-statement or a block-statement). Our goal was to aggregate execution; as such, we made design decisions to appropriately filter operations that were considered extraneous but that may be crucial to surface in other contexts.

\section{User Study}
We conducted an exploratory user study to understand how the workflow supported by \textsc{CrossCode} differs from existing tools, and if the representations proposed are useful to understand and communicate computer programs.

\begin{figure}[h]
  \includegraphics[width=0.5\textwidth]{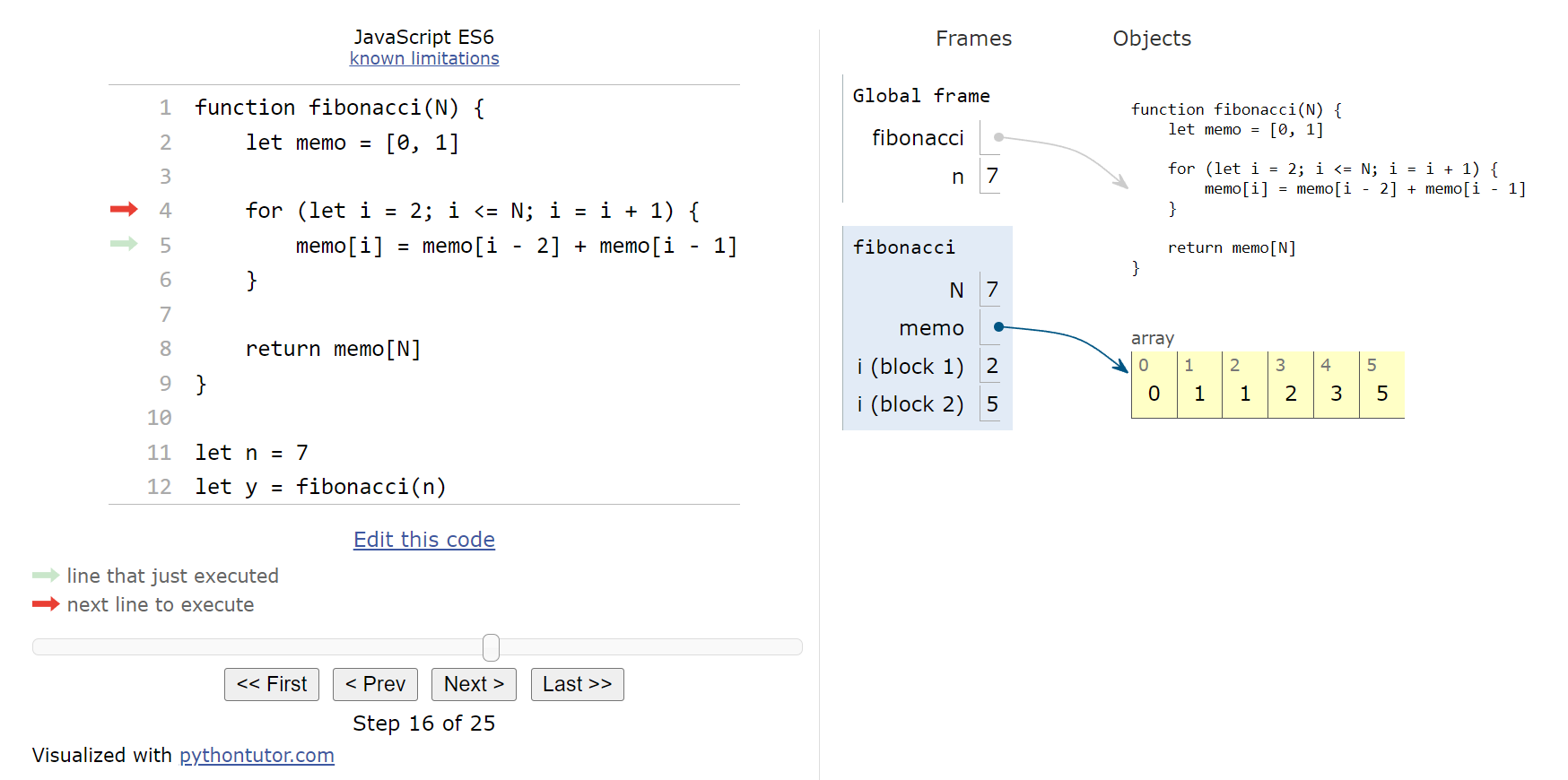}
  \caption{Screenshot of Python Tutor showing JavaScript code that iteratively computes the n-th number in the Fibonacci sequence. }
  \Description{A screenshot of Python Tutor web application. On left is shown JavaScript source code. Below the source code is a timeline of execution, with buttons to step forward and backward. On the right is a memory model graphic of the program state with boxes containing values and arrows as pointers.}
  \label{fig:pythontutor}
\end{figure}

We compared \textsc{CrossCode} against two other conditions in debugging tasks:

\begin{enumerate}
    \item \textsc{Python Tutor}, which is a widely online web app for visualizing program execution (Figure \ref{fig:pythontutor}) \cite{PythonTutor}. It was chosen as it is the state-of-the-art in program visualization tools and a line-by-line visualization. It uses a two-column layout, with source code on the left, and a depiction of the current execution state on the right. It includes a global timeline through which users can step to different parts of the execution, and buttons to go to the next or previous steps. Clicking on a line of source code allows setting a breakpoint, which is visually indicated along the timeline.
    \item \textsc{Drawing}, instructors and programmers frequently draw out program execution (e.g., on the blackboard) to communicate or reason about program behavior themselves. This condition was chosen as it enables creation of completely flexible representations, which can highlight deficiencies in the representation of \textsc{CrossCode} or \textsc{Python Tutor}. Participants were allowed to use pen and paper or a digital drawing tool.
\end{enumerate}

Our goal was not to compare the performance (i.e., the speed of completing the debugging task) but rather to elicit the representational affordances of each tool. The studies lasted between 90 minutes and 1 hour and 45 minutes each, and participants were compensated \$40 USD for their time.

\subsection{Participants}
We recruited six participants with extensive programming experience who work and/or teach in computer science and related fields (Table ~\ref{table:DemographicsTable}).

\begin{table}
\centering
    \caption{Participant demographics, experience in years. Teaching includes experience as a teaching assistant.}
    \begin{tabular}{cp{1.7cm}p{0.5cm}p{1cm}p{1.5cm}p{1.4cm}}
        \toprule
        ID & Title                      & Age  & Gender & Programming\newline Experience & Teaching\newline Experience \\ \midrule
        P1 & Assistant\newline Professor        & 33   & Male   & 10 & 3   \\ \hline
        P2 & PhD Student              & 30   & Male   & 10 & 3   \\ \hline
        P3 & Software\newline Engineer          & 26   & Male   & 14 & 1   \\ \hline
        P4 & PhD Student              & 27   & Male   & 8  & 2.5 \\ \hline
        P5 & Professor                  & 54   & Male   & 35 & 34  \\ \hline
        P6 & PhD Student              & 24   & Male   & 6  & 0.5 \\ \bottomrule
  \end{tabular}
  \label{table:DemographicsTable}
\end{table}

\subsection{Study Procedure}
\subsubsection{(5 minutes) Introduction} Participants were provided with a brief description of the research and study goal. It was emphasized that the goal of the debugging tasks was not to reach the solution as fast as possible but to use the problem to reflect on the usefulness of the tool and its representation.
\subsubsection{(15 minutes) Guided Walkthrough of \textsc{CrossCode}} Participants were presented with an introduction to the visualization techniques available within \textsc{CrossCode} and were guided through its various features. If participants were unfamiliar with \textsc{Python Tutor}, a brief tutorial on the use of the tool was provided.
\subsubsection{(10 minutes x 6) Debugging Tasks} Participants were presented with six debugging tasks, with the goal of localizing and fixing a bug. Each task had a single bug, which included one of the following: off-by-one array index, incorrect loop bound, and incorrect base case. Two debugging tasks were performed per condition (i.e., one iterative and one recursive). The ordering of the conditions and the task chosen were appropriately counter-balanced, however, the iteration task always occurred first followed by the recursive task. As mentioned above, the goal of these tasks was not to be challenging, but to serve as a sandbox to explore these tools. After each task, participants were asked to reflect on whether the representation of the tool was useful for their own understanding of the algorithm and if they were to explain this algorithm and their solution to a novice, whether this tool would be conducive to that explanation. In doing so, it could reveal conflicts between each condition and the user's own mental model. 

The six tasks involved debugging programs that:
        \begin{enumerate}
            \item {[Iterative]} \textit{Inserted into a sorted list such that it remains sorted.}
            \item {[Iterative]} \textit{Returned the n'th Fibonacci number.}
            \item {[Iterative]} \textit{Reversed a list in-place.}
            \item {[Recursive]} \textit{Returned sorted list using out-of-place quick-sort.}
            \item {[Recursive]} \textit{Returned sorted list using merge-sort.}
            \item {[Recursive]} \textit{Returned index of a value in a sorted list with binary search.}
        \end{enumerate}
\subsubsection{(15 minutes) Post-questionnaire and Interview} Participants were administered a questionnaire on the usability and usefulness of \textsc{CrossCode} through 7-point Likert Scale questions. During a semi-structured post-interview, participants were asked to indicate which tool they perceived to operate closer to the level that they understand code, explain how their workflow and approach to the problem differed between tools, and which tool would best facilitate communication.

\subsection{Study Limitations} 
This study has several notable limitations. The types of tasks were artificial and had far simpler bugs than those encountered in the wild, and as such, these results may not generalize to tasks outside typical algorithm procedures. There are notable sources of bias that make our findings subject to error: Participants likely knew \textsc{CrossCode} was created by the researchers, all participants were male, the sample size was modest at best, and the analysis was performed only by the primary author. As all participants are experts, it is not directly evident whether \textsc{CrossCode} helps novices build better mental models and learn effective debugging processes.

\section{Study Results}

The participants found \textsc{CrossCode}'s representation to be understandable and effective. With \textsc{CrossCode}, participants rated themselves to maintain a strong sense of their place in the execution (1/6 Strongly Agree; 3/6 Agree; 1/6 Slightly Agree; 1/6 Neutral), color encoding and traces were easy to understand (5/6 Agree; 1/6 Slightly Agree) and data animations were easy to follow (4/6 Strongly Agree; 1/6 Agree; 1/6 Slightly Agree).

We conducted a thematic analysis of participant reflections during their task and interview focusing on the role of \textsc{CrossCode} as a tool for (1) debugging, (2) program understanding (that is, developing valid mental models of the code), and (3) visual aid to explanations. The first author coded the transcript, deriving 48 codes (e.g., \textit{``Difficult to establish a convention with Paper \& Paper''}), which were synthesized into seven primary themes. Below, we report the results of the thematic analysis.

\subsection{CrossCode for Debugging}
CrossCode added context and continuity to debugging, which was perceived to be helpful in orienting oneself, as well as locating and diagnosing bugs.

\subsubsection{Control Flow View added necessary context to navigate the execution trace} 
We observed that instead of internalizing the control flow, participants exploited \textsc{CrossCode}'s contextual cues to maintain a sense of place, for example, since \textit{``CrossCode shows the history behind how it got there, I was able to go back in time''} (P2) unlike \textsc{Python Tutor} where \textit{``you don't have any knowledge of how we got there because it's just showing the current frame.''} P4 notes that this led to a slower debugging process:

\begin{quote}
    ``I didn't feel confident in my ability to remember where I'd been or what to look at [with \textsc{Python Tutor}]... I just had to absorb all the information at each step... I had to step back several times and go to the beginning because I didn't really remember anything in between.'' -P4
\end{quote}

Similarly, in \textsc{Python Tutor}, P1 \textit{``kind of scrubbed [through execution] randomly... I don't know where to focus on but rather just scrolling back and forth to try and spot if there is something wrong on this line.''} The \textit{Control Flow View} in \textsc{CrossCode} also enabled new workflows, such as allowing participants to visually deduce an erroneous program, e.g.,

\begin{quote}
``So, I've got 1,2,3,4 iterations before I fail... I can see that I've only got three items left [in the array] and so just from this representation I can say, `Why am I doing one more?' And of course I don't have that clue in any of the other tools... that requires some sophistication but that's huge.'' - P5
\end{quote}

\subsubsection{Animations, color encodings, and traces provided visible indicators of change} Animations enabled participants to follow the data flow of values. For example, P6 \textit{``would just go back and see the animation again like this [grabbing the control flow cursor]... and say, `Okay where did this [item in array] one go?' And I'll just slowly drag [the control flow cursor] and see''} (P6). Aggregations complemented these data views, P2 found it \textit{``quicker and faster [to debug] in \textsc{CrossCode} [compared to \textsc{Python Tutor}] because I was able to quickly understand what was being changed, or quickly go through the whole for-loop to find something.''} Unlike \textsc{Python Tutor}, which provided no indication of change, \textsc{CrossCode}'s color encodings for reads and writes helped infer patterns, e.g.,

\begin{quote}
``One of the great things about this animation here is that it shows which values are currently being modified, it shows that these two things [last two items of array] are being used... and based on those two things it's going to generate the next one [fibonacci number].'' -P2
\end{quote}

When \textsc{Drawing}, participants indicated a change by crossing out previous values, which although not scalable to larger inputs, served as an ad hoc mechanism to abstract over time and make predictions of program behavior. While \textsc{CrossCode}'s residuals also depicted past values, they were not sufficient when comparing values across substantially different points in time. P3 would have \textit{``liked to be able to have a comparison, I could say 'I think this frame where we are right now is really interesting' take a snapshot and then go to the end and see the comparison''} (P3).

\subsection{CrossCode for Program Understanding}
CrossCode served as a visual aid to build valid and useful mental of the code. Aggregations provided the ability to synthesize key concepts, while abbreviations and the \textit{Control Flow View} enabled navigation across repetitive structures and to key steps effectively.

\subsubsection{Overview and abbreviations of control flow allows opportunistic navigation to interesting steps} When \textsc{Drawing}, participants did not draw or simulate mental computations at every line, but instead opportunistically skipped repetitive steps (i.e. loops) or trivial steps (e.g. variable initialization). When investigating for-loops, participants frequently computed the first iteration of a loop manually, hypothesized a pattern, verified it over the next few iterations, and then skipped over subsequent iterations under all three conditions. \textsc{CrossCode} provided a structure for both forms of navigation. For P5, aggregations extended the functionality of breakpoints offered in traditional debuggers:

\begin{quote}
    ``I found it natural to navigate the function by looking into it. I like being able to treat these [sub-steps] atomically, in fact, these are a lot like breakpoints to me, except I don't have to set them manually.'' -P5
\end{quote}

\subsubsection{Aggregations alleviate the need to build mental models line-by-line} 
Participants found \textsc{CrossCode}'s ability to chunk operations into \textit{Steps} to be closer to their own mental model of code execution compared to stepping through the code line-by-line. For example, P1 shared their reasoning process as \textit{``you can group all these steps together by just saying this creates a left array, and that creates a right array... like what you had [in \textsc{CrossCode}].''} Similarly, when \textsc{Drawing}, P2, P3, and P4 explicitly drew chunks on top of the source code when tracing through the code by hand. P2 described their drawing process as:

\begin{quote}
    ``I divided different parts of the code... based on what they are supposed to do. So, I've some feelings about some chunks that are correct, that are correct [which P2 explicitly annotated with a checkmark], and I'm suspicious about other chunks.'' -P2
\end{quote}

\subsection{CrossCode as an Instructional Visual Aid}
CrossCode's aggregations were perceived as useful in facilitating explanation processes. However, visual aids are often specialized, and participants reported several opportunities to customize \textsc{CrossCode}'s representation to better direct explanations.

\subsubsection{Aggregations map to explanation processes better than line-by-line tools}  \textsc{CrossCode}'s aggregations enabled a gradual increase in complexity; P1 noted that aggregations better mapped to concepts, which usually span many lines of code:

\begin{quote}
``Line-by-line is the lowest level way to walk through what is going on... You wouldn't use that way to explain to people, but rather have concepts, like each concept maps to a function and then you create a link between those functions without relying on this [\textsc{Python Tutor}] linear representation.'' -P1
\end{quote}

P6 found \textsc{CrossCode} to be \textit{``the best in explaining, it's sort of like a pipeline, which visualized everything pretty well.''} \textsc{CrossCode} was also found to be applicable to meta-thinking exercises. For example, when working through an algorithm with a student, P5 would want to ask:

\begin{quote}
``Okay, if I were to execute this loop, what's the effect going to be like?'... and then if I treat it [the for-loop] as atomic it's easy to test that in the \textsc{CrossCode} tool... in \textsc{Python Tutor} I'd have to click next a bunch of times or set a breakpoint and then remember that I had set it.'' -P5
\end{quote}

P4 found the ability to unabbreviate loops critical to explanation processes, \textit{``for these loop-centric algorithms, of which there are many, it's really important to be able to roll and unroll loops... you want to be able to visualize them'' (P4).}

\subsubsection{Customizations are used for higher-order explanations} When \textsc{Drawing}, participants frequently annotated on existing variable values (e.g., annotating the index \verb | i| next to the item it indexes in the array) to both debug and explain code. P2 would have liked to \textit{``visually show what [index] is list[n - 1 - i] pointing at, and what [index] is list[i] pointing at. What [index] is temp pointing at?.''} Annotations can also help direct attention when teaching, \textit{``for the stuff I taught... like in data-structures, with rotating trees... you need to know where to look or you'll be lost... \textsc{CrossCode} is much closer to what I'd like, but I'd use paper or pen alongside it''} (P4).

\subsection{Drawbacks and Limitations of \textsc{CrossCode}}
The main overhead of \textsc{CrossCode} is in specifying the level of abstraction, which can slow debugging of simple examples compared to \textsc{Python Tutor}. While a user could have selected a part of the source code to navigate to the associated level of abstraction, it was scarcely used, and nearly all exploration was top-down. P4 and P3 also reported difficulties synchronizing between the \textit{Control Flow View} and the \textit{Data View}. And, P3, a practitioner, found it difficult to see the applicability of \textsc{CrossCode} as \textit{``98\% of the problems [they] have to explain to coworkers don't involve primitive data.''}

\subsubsection{Issues with Changes in Abstraction and Discoverability of the Appropriate Level of Abstraction} Participants reported several usability problems with \textsc{CrossCode}. P3 found that navigating the timeline in \textsc{Python Tutor} is straightforward and convenient, for example, \textit{``as long as I can kind of get there''} even with \textit{``trial and error, I don't feel that weird,''}  (P3) and P6 found \textit{``expanding and shrinking to be a little counterintuitive [in \textsc{CrossCode}].''} Selecting a piece of code to navigate to that level of abstraction was hard to discover and scarcely used. Adding computational suggestions to appropriate levels of abstraction may circumvent this: P2 found the resulting representation after \textit{``select[ing] this [base case], that it shows me [expanded function graph], that's good. But it's not clear. I would have... descriptions or indicators on it [the base case].''} P3 suggested that previewing abstractions could make switching between them more approachable, \textit{``Where you could hover over a ... checkbox [conditional check] it shows a literal representation of those numbers [being compared] at that moment''} (P3). \textsc{CrossCode}'s default level of abstraction can also be deceptive. P1, for example, did not realize that a function-call was initially recursive, \textit{``I just remember when I clicked that `left', it just automatically did all the recursion functions, so I said, oh `That's what it does.'''} P1-P5 also noted that they would need more time to become proficient and comfortable using \textsc{CrossCode} compared to \textsc{Python Tutor}, for example, \textit{``It's just that there are more features, I'd need more training''} (P5).

\subsubsection{Disconnect between the \textit{Control Flow View} and the \textit{Data View}} P4 noted difficulties in tracking changes in \textit{Control Flow View} and \textit{Data View} simultaneously:

\begin{quote}
``When I'm looking at [control flow] and unfolding stuff, I'm very much focused on it... the marble [cursor] is approaching solving the problem of what to look at. But there is nothing linking marble moving and the animations.'' -P4
\end{quote}

Both P4 and P3 wished that the data state be better integrated with the control flow, for example, P4 \textit{``appreciated like if you hover over this [control flow] it hovers over to the left [source code] as well, I wonder if it could hover over to the right [data].''}


\section{Discussion And Future Work}
By externalizing control flow, \textsc{CrossCode} allowed participants to maintain a sense of place in program execution, and the color encoding and animations provided strong indicators of change. \textsc{CrossCode} helped locate errors and was conducive to explanations. 

\subsection{Study Implications}
There are significant implications of the workflows enabled by \textsc{CrossCode} which are explored below.

\subsubsection{\textsc{CrossCode} encouraged a top-down program understanding strategy.}
\label{section:Workflows}
\textsc{CrossCode} enabled new workflows for debugging and explaining computer programs; however, the overall interaction strategy was also singular because participants explored program execution top-down. Even though participants could have selected source code to navigate to a specific level of abstraction, they rarely did, possibly because there was not an indicator to select code nor a description of what selecting a particular piece of code would do. There was also a sudden change in the control flow representation, so context was difficult to preserve when switching between varying abstractions. Previewing a level of abstraction before committing to it may help preserve context. In addition, adding more navigation strategies, such as by selecting specific data of interest, can also help trace the code (that is, enabling program slicing \cite{ProgramSlicing}).

\subsubsection{\textsc{CrossCode}'s idealized representation can complement a programmer's internal model.}
\label{section:IdealizedRepresentation}
\textsc{CrossCode}'s does not display an accurate or complete model of the underlying computation, and this may sometimes be deceptive. However, a programmer's internal model of computation does not match with the computer's exact semantics either. Programmers do not code for a machine, but rather for an idealized execution environment \cite{NotationalMachinesJuha}, commonly referred to as notional machines (i.e., classes of abstractions over a computer's exact semantics). Notional machines have been used as a pedagogical device to help learners build valid mental models of a program, sometimes manifesting themselves through diagrams of memory models and code execution \cite{EngageAgainstMachine}. The notional machine proposed by \textsc{CrossCode}, if given enough time learn, may help users develop more useful and consistent mental models of code than existing systems since it is closer to how program behavior is communicated and explained than line-by-line tools.

\subsubsection{\textsc{CrossCode}'s model of navigation may scale to complex programs.}
\label{section:Scalability}
Compared to existing line-by-line program visualization tools such as \textsc{Python Tutor}, \textsc{CrossCode} may have a greater potential to scale to more complex code as it aggregates operations rather than presenting them at a static level of detail. However, challenges remain in extending the current design to common programming paradigms such as object-oriented programming, which will require specialized abstractions (i.e., abstraction based on syntax tree will not solely capture models of classes and concepts such as inheritance) and representations (e.g., \cite{MemView}). Other considerations, such as scaling across multiple files or to more complex states, will require specialized mechanisms for managing complexity in the \textit{Control Flow View} and the \textit{Data View}. Regardless, we believe that the guiding principle behind \textsc{CrossCode}, i.e., designing representations of execution based on how program behavior is visually explained, generalizes to these structures as well.


\subsection{Future Work}
The evaluation points to several critical directions for the design of future programming interfaces. Here, we outline three key areas of investigation for future work.

\subsubsection{Aggregations beyond syntax.}
Our results suggest that top-down exploration of execution based on the program's syntax tree facilitates new and potentially more efficient program understanding and debugging processes. However, there are other appropriate abstraction strategies that were not explored. For one, \textsc{CrossCode} treats all source code the same: but there is a rich set of semantics in how the users write the source code and their documentation, e.g., users may implicitly chunk lines of code that serve a similar function. Programs are a form of human communication, akin to natural language: \textit{Can statistical properties of the code (i.e., \cite{BigCode}) be leveraged to infer appropriate abstraction strategies?} The data flow may also be used to inform potential abstractions, for example, a loop that divides a list into two sub-lists, can be described as two aggregate steps: filling in the first sub-list and then filling in the second sub-list, a description which cannot be captured by \textsc{CrossCode}'s aggregations.


\subsubsection{Bridging the control flow and data representations.}
\textsc{CrossCode} separates a program's control flow from its data state. While doing so provides flexibility in the placement, participants found it difficult to track and synchronize between the two views. Various techniques from prior work in live programming and information visualization can help bridge the two views. For one, data can be visually annotated next to control flow structures, e.g., using similar visual techniques as Projection Boxes \cite{ProjectionBoxes}. Control flow may be navigated by data (e.g., clicking on an item in the list could navigate execution to when that item was last modified). The same abbreviations that occur in the control flow can be extended to the data encoding: If the iterations of a loop are abbreviated, then the corresponding list items can be appropriately abbreviated in the data representation.

\subsubsection{Externalizing control flow of synchronous executions and multi-threaded programs.}
\label{section:Asynchrony}
Current representations of control flow in \textsc{CrossCode} support a single notion of time. Designing representations for synchronous operations such as web API calls or file I/O could help manage and communicate classes of bugs such as race conditions which are often difficult to form an accurate mental model of. This is especially true in distributed systems, where complex and time-sensitive operations can be difficult to diagnose with traditional debugging tools. The ability to communicate the behavior of concurrent systems to others could facilitate collaboration and improve the overall quality of instruction of distributed systems. By analyzing visual organizations and metaphors used to teach concurrency, future research can inform the design of appropriate representations of concepts such as shared resources, ownership, and collisions, which are otherwise difficult to surface.

\section{Conclusion}
The main conclusion from this research is that aggregation of a program's execution based on the syntax tree can better support program debugging and explanation workflows than line-by-line debuggers. Through a formative study of hand-designed depictions of program execution, we synthesized three design patterns: visualizations should aggregate execution, abbreviate repetitive operations, and present an overview of the execution space. We realized these patterns in \textsc{CrossCode}, a program visualization system capable of visualizing and navigating between multiple levels of abstractions. The results of an expert evaluation comparing \textsc{CrossCode}, \textsc{Python Tutor}, and \textsc{Drawing} highlighted the usefulness of \textsc{CrossCode}'s representation of control flow to maintain context, as well as its use of aggregations to explore execution beyond line-by-line methods.

\begin{acks}
We thank the anonymous reviewers for their valuable and constructive feedback, Matthew Beaudouin-Lafon for insightful discussions, and members of the Creativity Lab for their feedback and generous assistance.
\end{acks}

\bibliographystyle{ACM-Reference-Format}
\bibliography{base}


\end{document}